\documentclass[prl,aps,longbibliography,floats,superscriptaddress,twocolumn]{revtex4-1}

\usepackage[utf8]{inputenc}
\usepackage{amsmath,amssymb,amsthm,amsbsy,bm}
\usepackage{dsfont,verbatim,varwidth,dcolumn}
\usepackage{graphicx,epsfig,epstopdf,subfigure}
\usepackage[usenames,dvipsnames]{xcolor}
\usepackage{color,soul,subfiles}
\usepackage[colorlinks]{hyperref}

\newcommand{\gt}{g_{\text{2-ter}}}

\makeatother

\begin{document}

\title{Emergence of Neutral Modes in Laughlin-like Fractional Quantum Hall Phases}

\author{Udit Khanna}
\altaffiliation{Present Address: Department of Physics, Bar-Ilan University, Ramat-Gan 52900, Israel}
\email{uditkhanna10@gmail.com}
\affiliation{Raymond and Beverly Sackler School of Physics and Astronomy, Tel-Aviv University, Tel Aviv, 6997801, Israel}
\affiliation{Department of Condensed Matter Physics, Weizmann Institute of Science, Rehovot 76100, Israel}

\author{Moshe Goldstein}
\affiliation{Raymond and Beverly Sackler School of Physics and Astronomy, Tel-Aviv University, Tel Aviv, 6997801, Israel}

\author{Yuval Gefen}
\affiliation{Department of Condensed Matter Physics, Weizmann Institute of Science, Rehovot 76100, Israel}

\begin{abstract}
Chiral gapless boundary modes are characteristic of quantum Hall (QH) states. For hole-conjugate fractional QH phases 
counterpropagating edge modes (upstream and downstream) are expected. In the presence of electrostatic interactions 
and disorder these modes may renormalize into charge and upstream neutral modes. Orthodox models of Laughlin phases 
anticipate only a downstream charge mode. Here we show that in the latter case, in the presence of a smooth 
confining potential, edge reconstruction leads to the emergence of pairs of counterpropagating modes, which, by 
way of mode renormalization, may give rise to nontopological upstream neutral modes, possessing nontrivial statistics. 
This may explain the experimental observation of ubiquitous neutral modes, and the overwhelming suppression of anyonic 
interference in Mach-Zehnder interferometry platforms. We also point out other signatures of such edge reconstruction. 
\end{abstract}

\maketitle

{\it Introduction.}--Transport properties of two-dimensional topological insulators, such as, quantum Hall (QH) states are determined 
by gapless edge modes~\cite{Halperin1982}. The structure of the boundary is, in turn, constrained by the bulk 
topological invariants~\cite{WenBook}. 
Particlelike fractional QH states (described by a positive definite $K$ matrix) are expected
to support one or more gapless ``downstream'' chiral edge modes~\cite{WenBook,Wen1990,Wen1992}.
By contrast, hole-conjugate states host multiple branches of boundary modes, some of which propagate upstream, 
thus satisfying bulk topological constraints~\cite{Wen91A,Wen91B,MacDonald_PRL_90,Johnson_PRL_91}. 
Such counterpropagating edge modes are renormalized by disorder-induced tunneling
and intermode interactions, which may lead to the emergence of upstream neutral modes~\cite{KFP1994,KaneFisher_PRB_95}. 
Notwithstanding the different classes of topological bulk states, neutral modes appear to be ubiquitous. 
Experimental signatures of the latter include upstream heat transport with net zero charge~\cite{KaneFisher_PRB_97} as well as 
suppression of anyonic interference~\cite{Moshe_PRL_2016}. These 
have been observed in hole-conjugate and non-Abelian QH states~\cite{Yacoby2012,Bid2009,Bid2010,
Gurman2012,Yaron2012,Inoue2014,Heiblum2019,Sabo2017,Heiblum2021,Ballistic_Heatflow}, and most surprisingly, 
in particlelike states as well~\cite{Inoue2014,Heiblum2019}.  

Laughlin states ($\nu = 1/m$ for odd $m$), the simplest example of particlelike phases, are expected 
to support a single downstream edge mode (hence no upstream neutral). 
However, transport measurements of these states~\cite{Inoue2014,Heiblum2019}
reveal that the structure of the edge is much more intricate. Specifically, Ref.~\cite{Inoue2014} observed that 
partial transmission of charge current through a quantum point contact (QPC) is accompanied by upstream electric 
noise (with no net current). Reference~\cite{Heiblum2019} observed that the visibility of the interference pattern in 
an electronic Mach-Zehnder interferometer decreases as the filling factor ($\nu$) is reduced from $2$ to $1$, and is 
fully suppressed for $\nu \leq 1$. For Laughlin states, these results are clearly inconsistent with the orthodox edge model, 
indicating the presence of additional neutral modes at the edge. The emergence of such modes at the QH edge has 
far-reaching implications. Upstream neutral modes may act as {\it which-path} detectors, thus suppressing anyonic 
interference signatures~\cite{Moshe_PRL_2016}. Furthermore, upstream neutrals may lead to the generation of shot 
noise with universal Fano factor {\it on a QPC conductance plateau}~\cite{Spanslatt2020,Jinhong2020}. Given the 
recent successes in observing of anyonic interference~\cite{Manfra2020,Kundu22} and measuring universal Fano 
factors~\cite{Biswas2021}, understanding the ubiquitous emergence of neutral modes, even at the edge of Laughlin 
phases, is of central importance.

A smooth confining potential at the boundary is known to induce quantum phase transitions at the edge, which leave the bulk unperturbed, 
in both integer~\cite{CSG1992,Dempsey1993,ChamonWen,Sondhi_PRL_96,FrancoBrey97,KunYangIQHS,Switching2017,Ganpathy21,Karmakar2020,IQHS2020} and 
fractional~\cite{Meir93,MacDonald_JP_93,KunYang03,KunYang_2002,KunYang_2003,KunYang_2008,KunYang_2009,
Ganpathy_PRB_03,Jain2014,Yang2021,Liangdong21} QH phases, as well as in time-reversal-invariant topological 
insulators~\cite{Yuval2017,Rosenow2021}. 
Such transitions (a.k.a.~{\it edge reconstruction}), which may lead to a change in the number, ordering, and/or the nature of the 
edge modes, are driven by the competition between the electrostatic effects of a smooth confining potential and the exchange 
and/or correlation energies of an incompressible QH state. For sufficiently smooth potentials, this competition leads to nucleation of additional
electronic strips (in QH phases) along the edge~\cite{Beltram2012,Thomas2014}, which define pairs of counterpropagating 
chiral modes at their respective boundaries. Hence, the structure of the reconstructed edge is not uniquely determined by the 
bulk-boundary correspondence. Specifically, the $K$ matrix entering the effective (1+1 D) boundary theory is no longer 
identical to the most compact $K$ matrix describing the bulk topological order. 
Anomalous bulk-boundary correspondence has also been proposed in other topological media~\cite{BrokenBulkEdge}. 
Edge reconstruction has several experimental manifestations, e.g., a quantized heat conductance much larger than 
expected from the orthodox edge structure~\cite{Anindya2022}, and the breakdown of quantization of tunneling exponents 
in the fractional QH regime~\cite{Exponents}. Additionally, intermode interactions and disorder-induced tunneling among 
these additional and the original (topological) edge modes may lead to a subsequent renormalization, which qualitatively modifies 
their nature and may even give rise to additional (nontopological) upstream neutral modes~\cite{WMG_PRL_2013}.

\begin{figure}[t]
  \centering
  \includegraphics[width=\columnwidth]{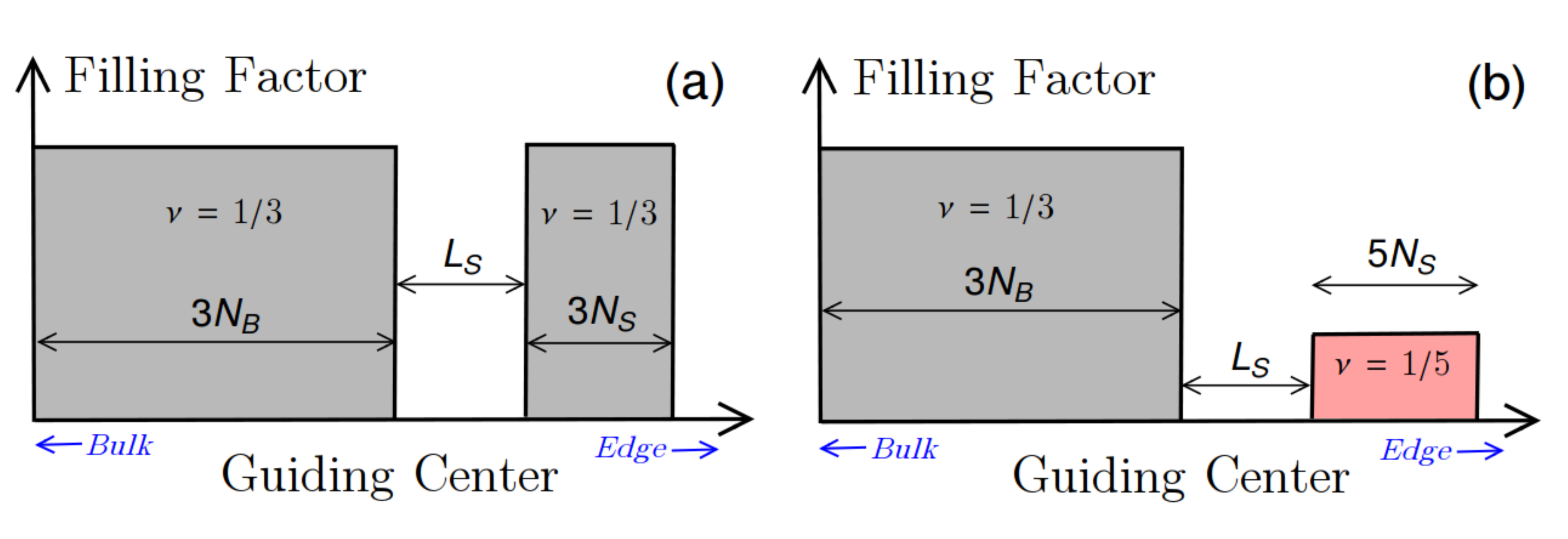}
  \caption{ Schematic of two {\it a priori} possible configurations at the edge of the bulk $\nu = 1/3$ phase. 
  For a sharp confining potential, there is a single ($\nu = 1/3$) quantum Hall droplet. 
  For smoother edge potentials, an additional side strip, with filling factor (a) $1/3$ or (b) $1/5$,  
  composed of $N_{S}$ electrons may be nucleated along the edge. The side strip is separated from the bulk 
  (comprising $N_{B}$ electrons) by $L_{S}$ guiding centers. }
\end{figure}

\begin{figure*}[t]
  \centering
  \includegraphics[width=0.24\textwidth]{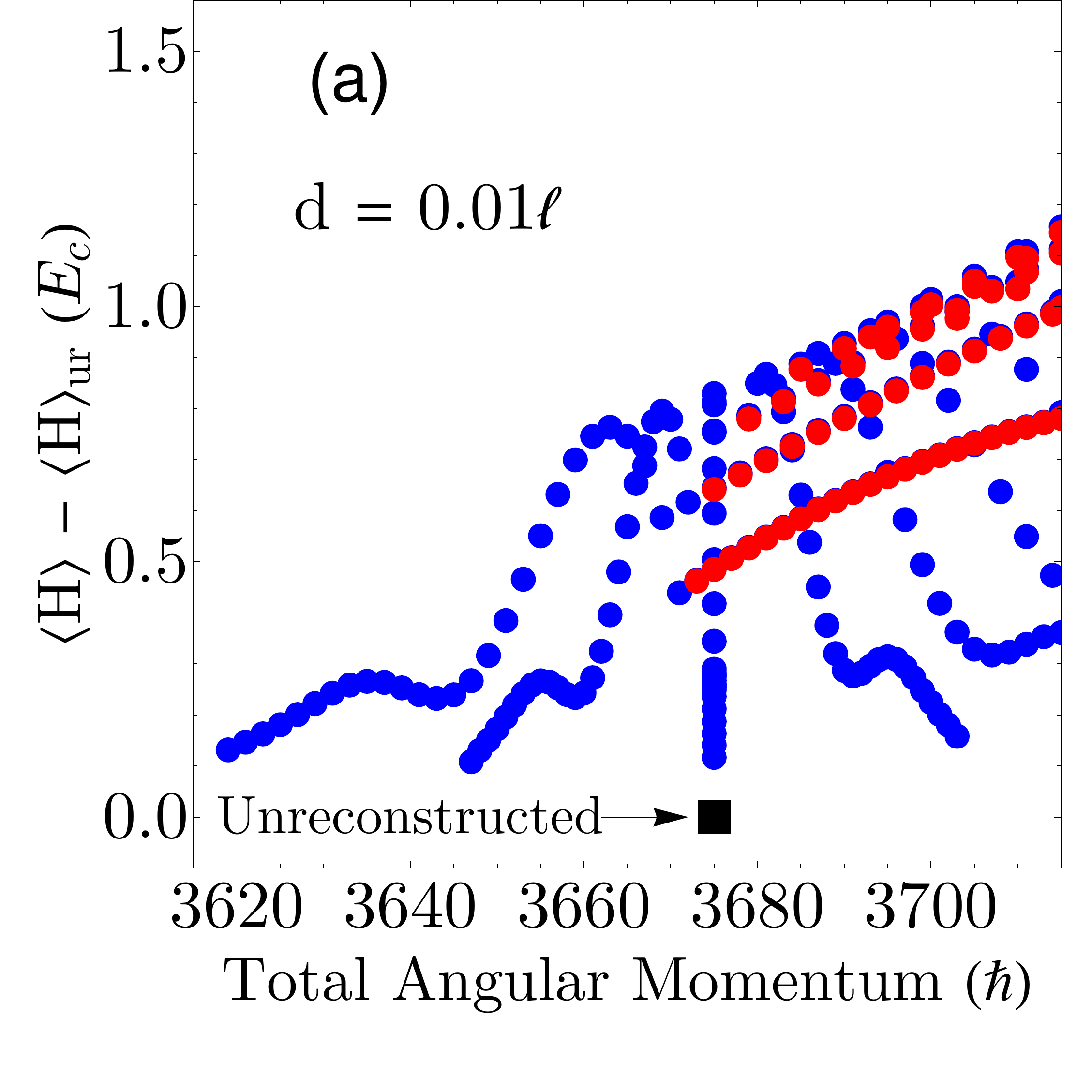}
  \includegraphics[width=0.24\textwidth]{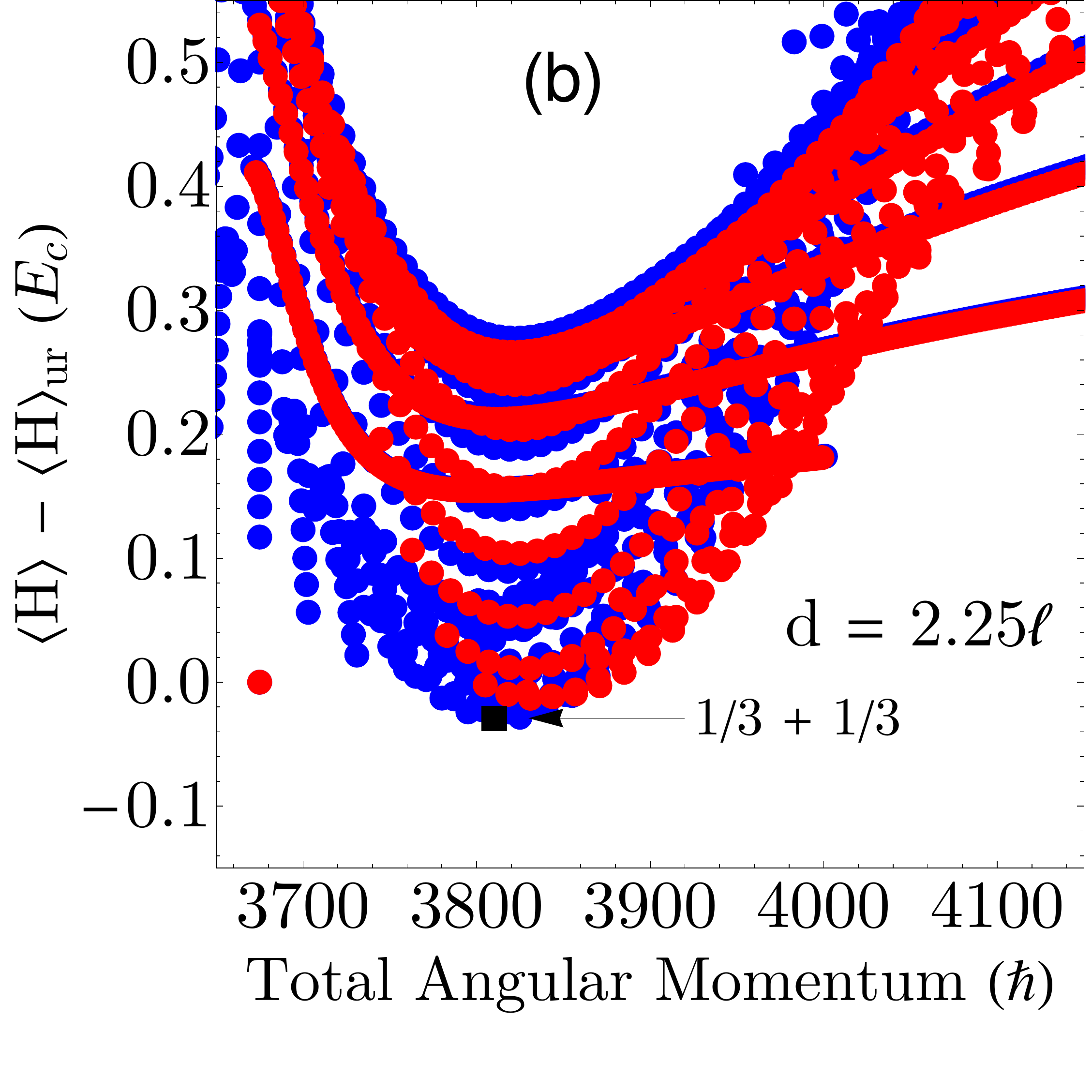}
  \includegraphics[width=0.24\textwidth]{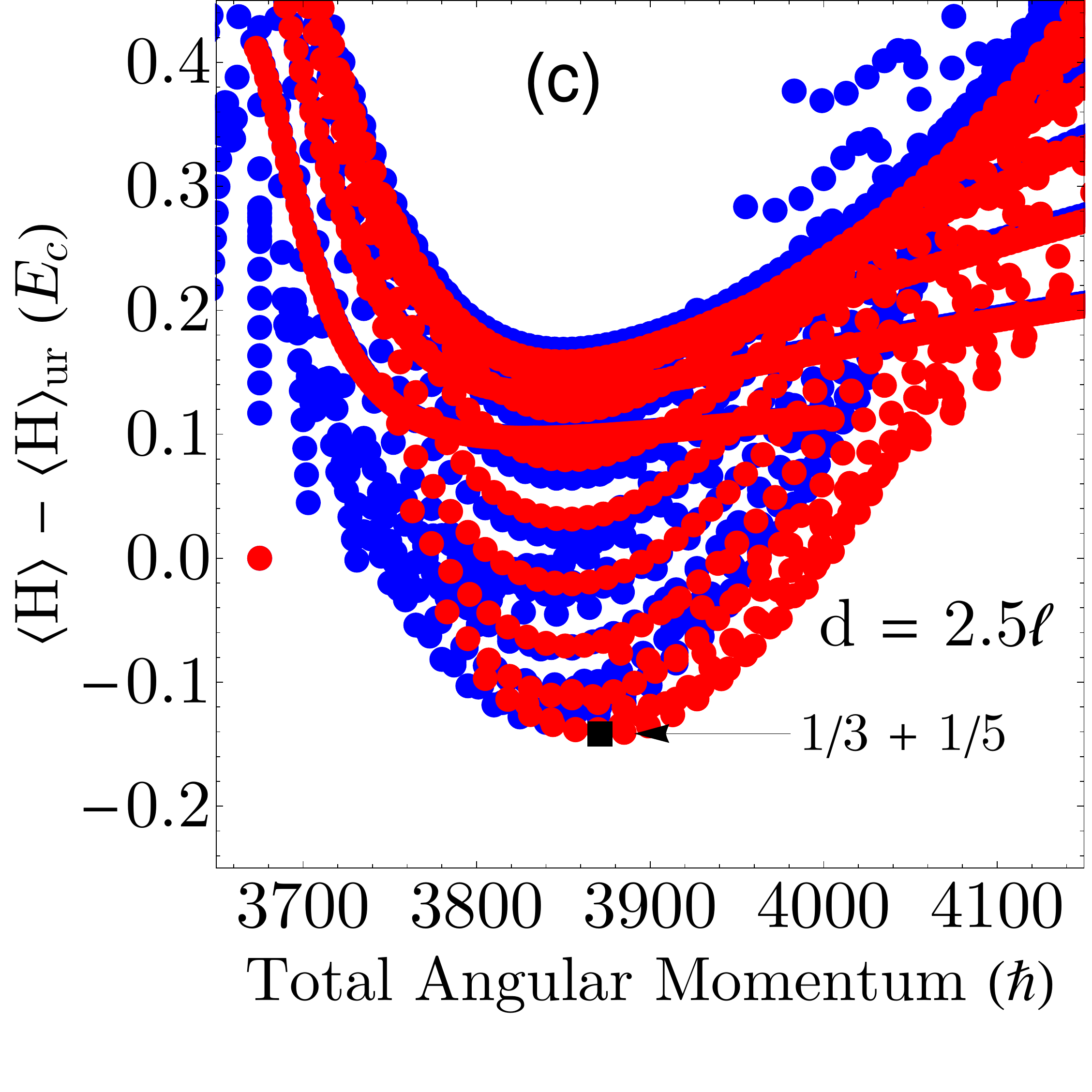}
  \includegraphics[width=0.24\textwidth]{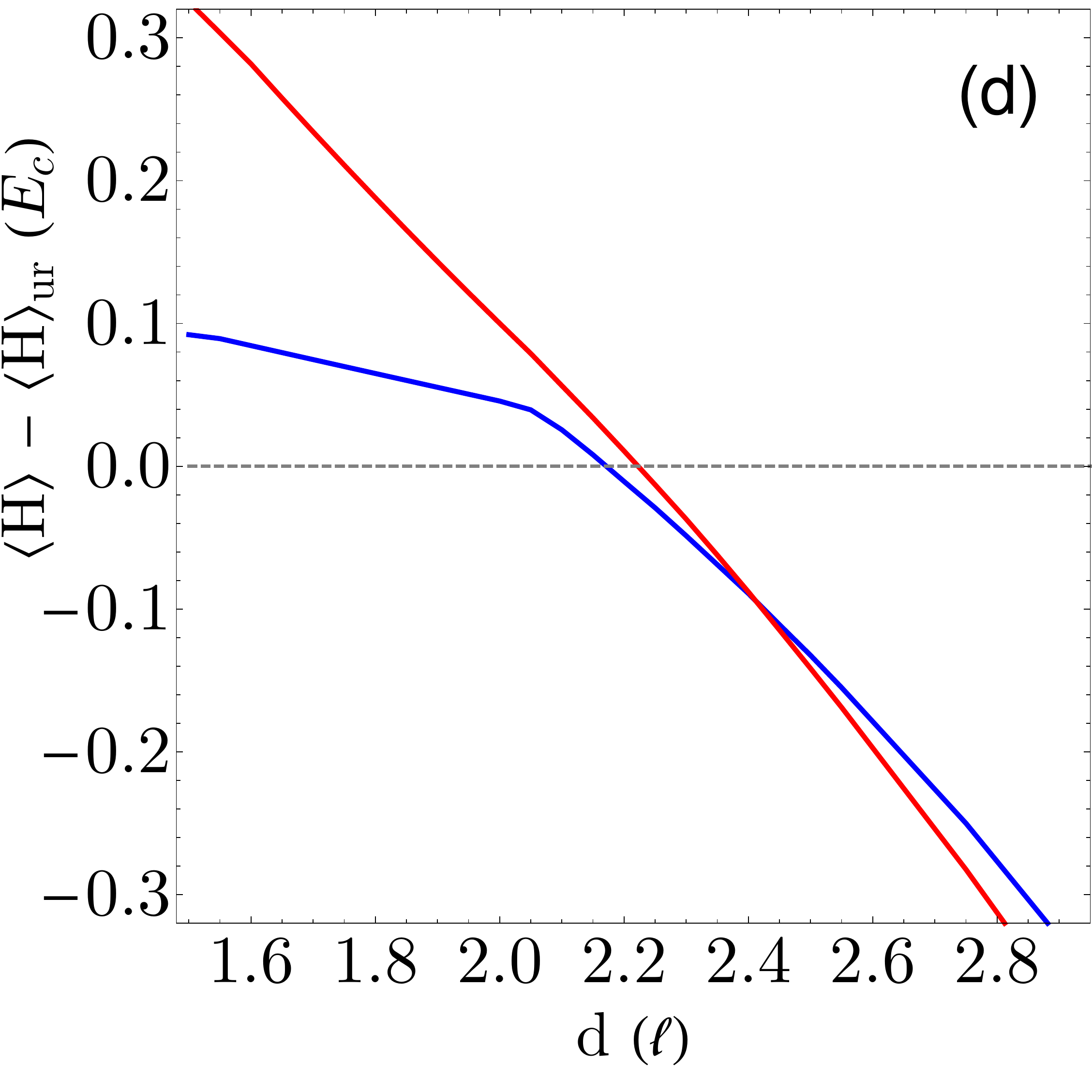}
  \caption{ Results of the variational calculations with 50 electrons. 
  (a)-(c) The energy ($\langle H \rangle$) of the 
  states in the two variational classes as a function of the 
  total angular momentum at (a) sharp ($d = 0.01 \ell$), (b) 
  moderately smooth ($d = 2.25 \ell$), and (c) very smooth 
  ($d = 2.50 \ell$) confining potentials, where $\ell$ is the magnetic length.
  In all cases, the energy of the unreconstructed state ($\langle H \rangle_{\text{ur}}$) has been
  subtracted. The blue (red) circles show energy of states with a side strip of $\nu = 1/3$ ($\nu = 1/5$). 
  The black square marks the state with the 
  lowest energy. (d) The lowest possible energy in the two variational classes 
  as a function of the smoothness of the confining potential (parameterized by $d/\ell$). The blue 
  (red) line corresponds to states with a side strip of $\nu = 1/3$ ($\nu = 1/5$). 
  For sharp edges the ground state is the unreconstructed $\nu = 1/3$ state 
  with angular momentum $3675 \hbar$. This state supports a single 
  chiral $e/3$ mode. For moderately smooth potentials ($ 2.17 < d/\ell < 2.42$), an additional $\nu = 1/3$ side strip  
  is nucleated, which defines a pair of counterpropagating $e/3$ modes (in addition to the chiral $e/3$ mode 
  arising from the bulk). For very smooth potentials ($d > 2.42 \ell$), a $\nu = 1/5$ side strip is generated, which 
  supports a pair of $e/5$ edge modes. }
\end{figure*}

Our challenge here is to account theoretically for the reconstruction and, subsequently, renormalization, 
of the edge of a Laughlin state (specifically, $\nu=1/3$). To reach this goal, we need to determine the precise filling factor 
of the additional side strip nucleated at a smooth edge.   
Figure~1 depicts two {\it a priori} possible edge configurations, which are considered in our analysis~\cite{fnoteDMRG}.
We stress that there is a qualitative difference between these two edge structures. 
The additional side strip of filling factor $1/3$ ($1/5$) defines counterpropagating modes of 
charge $e/3$ ($e/5$). For the case of $1/3$ side strip [Fig.~1(a)], subsequent renormalization of the modes 
(due to disorder-induced tunneling) would lead to localization of a pair of counterpropagating modes and may 
render transport experiments blind to the presence of reconstruction. On the other hand, for the $1/5$ strip [Fig.~1(b)], 
subsequent renormalization of the original $e/3$ mode and the upstream $e/5$ mode would not induce localization, 
and (as we demonstrate here) would have clear experimental manifestations. 
Our analysis identifies which structure is energetically preferable in a given parameter range.  

Exact diagonalization~\cite{KunYang_2002,KunYang_2003,KunYang_2008,KunYang_2009}, being limited to small
system sizes, does not allow us to obtain a quantized filling factor at the edge. 
We stress that such an analysis cannot clearly resolve the precise configuration of the
edge, and, in particular, is unable to predict whether upstream neutral modes do or do not emerge upon reconstruction.  
For this reason, we employ a variational analysis to study the edge~\cite{Meir93,IQHS2020}, which overcomes
these size limitations, while fully accounting for quantum correlations, inherently 
present in the Laughlin state. Specifically, we treat the strip-size ($N_{S}$) and separation ($L_{S}$) (cf. Fig.~1) as variational
parameters, and evaluate the energy of the states in both configurations 
as a function of the confining potential slope. 
When the confining potential is sharp, there is no edge reconstruction, i.e., the lowest
energy structure corresponds to $N_{S} = 0$. 
As shown in Fig.~2, for moderately smooth potentials, an additional side strip comprising a $\nu = 1/3$ phase emerges,
while for even smoother potentials, the filling factor of the side strip becomes $\nu = 1/5$. 
The edge modes of the latter structure, and their ensuing renormalization leading to the emergence of neutral modes, 
may account for the experimental results reported in Refs.~\cite{Inoue2014,Heiblum2019}.

Our results for the reconstructed edge structure may be verified in carefully designed transport experiments. 
Consider, for instance, the behavior of the two-terminal conductance ($\gt$) as a function of the sample length.
With a single gapless mode, $\gt$ is independent of the sample length and is determined solely by the
bulk filling factor (in this case $\gt = 1/3 \times e^{2}/h$). For reconstructed edges, though, 
the conductance may vary with the sample length~\cite{Gornyi21} due to intermode equilibration facilitated by interactions and 
disorder-induced tunneling~\cite{Yuval_AP_2017,Nosiglia2018}. 
For sufficiently long samples (with full edge equilibration), $\gt = 1/3 \times e^{2}/h$ for any edge
structure. For shorter samples (with no intermode equilibration), $\gt$ assumes the value 
$1 \times e^2/h$ ($11/15 \times e^2/h$) for a side strip of filling factor 1/3 (1/5).   
Furthermore, for the configuration with a $\nu = 1/5$ side strip, disorder-induced random tunneling and intermode interactions 
between the counterpropagating $e/3$ and $e/5$ modes lead to the emergence of new effective modes (Fig.~3), which, 
upon renormalization, may comprise upstream neutral modes. 
The experimental consequences of the emergence of such nontopological neutrals are similar to those discussed above for the 
case of hole-conjugate QH states.

\textit{Model.}--We analyze the edge of the $\nu = 1/3$ state in the disk geometry.  
The Hilbert space is restricted to the lowest Landau level and we assume spin-polarized electrons. 
In this limit, the bulk $\nu = 1/3$ state is well described by the Laughlin wave function~\cite{Laughlin83,JainCF} 
\begin{align}
  \Psi_{\frac{1}{3}, N} = \prod_{i=1}^{N} \bigg[ \prod_{j > i} \big( z_i - z_j \big)^{3} \bigg] e^{-\frac{1}{4} \sum_{i} |z_{i}|^2}, 
\end{align}
where $N$ is the number of electrons, $z_{j} = (x_{j} - i y_{j})/\ell$ is the position of $j{\text{th}}$ electron
and $\ell$ is the magnetic length. 
The Hamiltonian of the system is $H = H_{\text{ee}} + H_{\text{c}}$, where $H_{\text{ee}}$ is the   
electronic repulsion and $H_{\text{c}}$ is the confining potential (assumed to be circularly symmetric). 
Since $H$ is rotationally invariant, the many-body variational states
may be classified using the total angular momentum. 

We assume the electrons interact via the long-range Coulomb interaction 
($\frac{e^{2}}{4 \pi \epsilon_{0}} \sum_{i \neq j} 1/|\vec{r}_{i} - \vec{r}_j|$). 
The confining potential is modeled as the electrostatic potential of a positively charged background disk separated from the electron
gas by a distance $d$ along the magnetic field~\cite{KunYang_2002,KunYang_2003}. 
The density and radius of the background disk are fixed such that the full system is charge neutral~\cite{Supplemental}.
The slope of the ensuing potential is controlled by $d/\ell$, which is our tuning parameter.  
The potential is quite sharp for $d \sim 0$, and becomes smoother as $d$ increases. 
In our model $E_{\text{c}} = \frac{e^{2}}{4 \pi \epsilon_{0} \ell}$ sets the energy scale for both the electronic repulsion and  
the confining potential, and hence drops out of the analysis.

\textit{Variational analysis.}--Figure 1 shows the two classes of variational states considered here to describe the reconstructed edge of a 
$\nu = 1/3$ Laughlin state. Both classes represent product states  
($|\Psi_{\frac{1}{3}, N_{B}} \rangle \otimes |\Psi_{\frac{1}{m_{S}}, N_{S}, M_{S}} \rangle$) 
of the bulk and a single edge strip. The edge strip comprising $N_{S}$ electrons is described by a 
$\nu = 1/m_{S}$ Laughlin state ($m_{S} = 3, 5$) with $M_{S}$ quasiholes at the center. The corresponding 
(unnormalized) wave function is~\cite{Laughlin83,JainCF} 
\begin{align}
  \Psi_{\frac{1}{m_{S}}, N_{S}, M_{S}} = \prod_{i=1}^{N_{S}} \bigg[ z_{i}^{M_{S}} \bigg] 
  \bigg[ \prod_{j > i} \big( z_i - z_j \big)^{m_{S}} \bigg] e^{-\frac{1}{4} \sum_{i} |z_{i}|^2}. 
\end{align}
In our analysis, the total number of electrons ($N_{B} + N_{S}$) is fixed (to be 50 here). The number of electrons
in the strip ($N_{S}$) and the number of unoccupied guiding centers between the bulk and the strip ($L_{S} = M_{S} + 2 - 3N_{B}$)
are the two parameters that label the states in both the classes considered here. 
The energy ($\langle H \rangle$) of these states may be evaluated as a function of $d$, using 
standard classical Monte Carlo techniques~\cite{Supplemental}.
The unreconstructed state (without an additional edge strip) is included in both classes (corresponding to $N_{S} = 0$). 
It is the lowest energy state for sharp confining potentials. By contrast, the ground state supports an additional
edge strip (finite $N_{S}$ and $L_{S}$) for smoother potentials. 
The precise filling factor of this strip (and the nature 
of the additional counterpropagating modes) may be determined by comparing the energies of the states in the two classes.

\textit{Results.}--Figure 2 depicts the energy ($\langle H \rangle$) of the states in both classes, labeled by the total
angular momentum, for several values of $d$, which controls the sharpness of the confining potential. The blue (red) dots [in Figs.~2(a)-(c)] 
correspond to edges with a $\nu = 1/3$ ($\nu = 1/5$) side strip. The black square marks the lowest 
energy state. In all cases, the energy of the unreconstructed state was subtracted to  
ease the comparison. For a sharp confining potential [$d \lessapprox 2.1\ell$, Fig.~2(a)] the 
Laughlin state (with no additional side strip), supporting a single chiral $e/3$ edge mode, has the lowest energy (as expected).

The lowest energy state at smoother potentials [$d \gtrapprox 2.1\ell$] comprises an 
additional side strip. This side strip may have filling factor $1/3$ [Fig.~2(b)] for a moderately smooth  
potential ($N_{S}=15, L_{S}=11 \text{ for } d=2.25\ell$) or $1/5$ [Fig.~2(c)] for 
very shallow potential ($N_{S}=14, L_{S}=3 \text{ for } d=2.5\ell$). 
Figure 2(d) shows the variation of the lowest possible energy in the two classes with the 
confining potential slope. Evidently, the side strip filling factor is $1/3$ in the range 
$2.17 \ell < d < 2.42 \ell$, and switches to $1/5$ for larger $d$. 
The reconstructed edge configurations support, in addition to the single $e/3$ mode arising from the bulk, a pair of 
counterpropagating $e/3$ or $e/5$ modes. 
In the rest of this Letter, we focus on the experimental manifestations of these additional counterpropagating modes.

\textit{Transport signatures--two terminal conductance.}--The various edge structures obtained in our numerical 
analysis may be identified through their unique signatures 
in designed transport experiments. The (electric) two terminal conductance ($\gt$) as a function of the length of the
edge ($L$) is one such measurement. There, in the absence of edge equilibration, 
the chiral channels exiting the source contact are biased with respect to the modes entering it. 
The presence of impurities and potential disorder generates random tunneling between the (co- and counterpropagating) modes 
at the edge, which may facilitate intermode equilibration over a characteristic length $\ell_{\text{eq}}$. 
For $L \gg \ell_{\text{eq}}$, the two-terminal conductance is
$\gt = 1/3 \times e^{2}/h$ irrespective of the confining potential slope, reflecting the bulk filling factor.

More interesting is the $L \ll \ell_{\text{eq}}$ regime, where $\gt$ is sensitive 
to the detailed structure of the edge. For the unreconstructed edge, 
$\gt = 1/3 \times e^{2}/h$ for all values of $L$; the presence of a single chiral $1/3$ mode implies that the 
notion of equilibration is irrelevant. For reconstructed edges the additional pair of 
counterpropagating modes also contribute to $\gt$. For an edge comprising a $\nu = 1/3$ side stripe, 
$\gt = 1 \times e^{2}/h$ ($1/3 \times 3$). For a $\nu = 1/5$ side strip, $\gt = 
11/15 \times e^{2}/h$ ($1/3 + 2 \times 1/5$). Such unequilibrated
counterpropagating modes have been reported for other
bulk filling fractions~\cite{Lafont2019}.

\begin{figure}[t]
  \centering
  \includegraphics[width=\columnwidth]{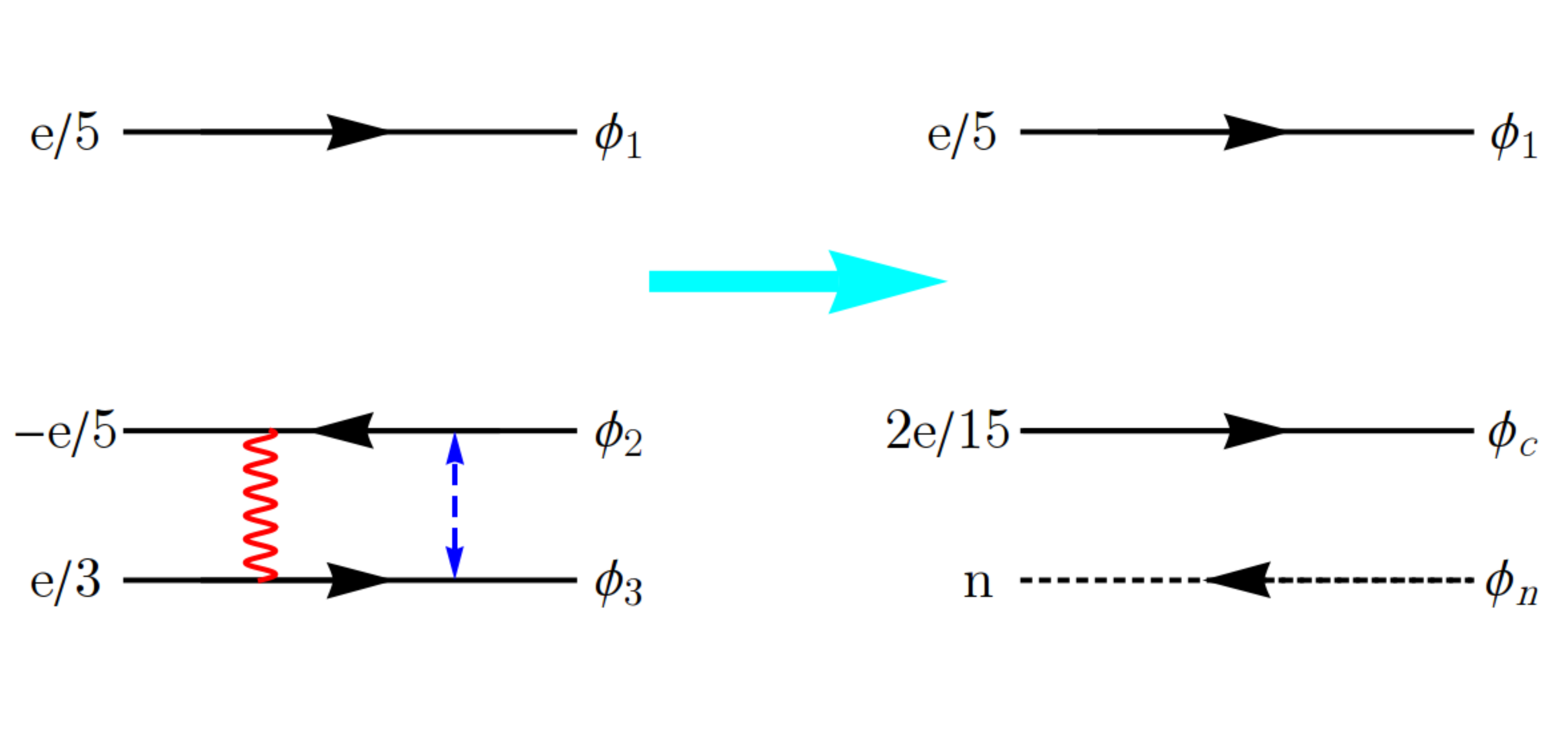}
  \caption{ For the edge structure with a $\nu = 1/5$ strip, 
  interactions (represented by the red curve) and disorder-induced electron tunneling
  (blue dashed line) between the inner two edge modes may lead to 
  emergence of a renormalized downstream charge ($\phi_{c}$) and an upstream neutral mode
  ($\phi_{n}$). The outermost mode is assumed to be completely decoupled from the inner two modes.  
  This idealization is justified by the variational analysis, which shows that 
  as the confining potential becomes shallower, the width of the side strip (proportional to $N_{S}$)
  increases faster than the separation of the strip ($L_{S}$). }
\end{figure}

\textit{Plateaus in conductance through a QPC.}--The existence of counterpropagating $1/5$ modes at the edge of a bulk $1/3$
state may also be detected through measurement of the
conductance across a QPC. For a $\nu = 1/3$ bulk phase,  
the conductance through a fully open QPC (transmission = 1) is expected to be $1/3 \times e^{2}/h$ 
(assuming full electrical equilibration). 
If the edge comprises an additional $\nu = 1/5$ strip and in the absence of strong edge renormalization, one may pinch-off the QPC 
such that the innermost $1/3$ and the upstream $1/5$ modes are fully reflected, and only the 
outermost $1/5$ mode is transmitted. Then a conductance plateau at $1/5 \times e^{2}/h$ is expected. 
As the bulk filling factor does not
deviate while tuning the QPC, this would be a smoking-gun signature of the presence of a 
$1/5$ strip at the edge of a bulk $1/3$ phase.

\textit{Neutral modes.}--Consider the reconstructed edge with a $\nu = 1/5$ side strip.  
The low energy dynamics of the three chiral modes may be described by (chiral) bosonic fields $\phi_{j}$ for $j = 1, 2, 3$ (outermost
being 1 and the innermost being 3)~\cite{WenBook,Wen1990,Wen1992}. The fields satisfy the  
Kac-Moody algebra, $[\phi_{j_{1}} (x), \phi_{j_{2}} (x^{\prime})] = i \pi \delta_{j_{1} j_{2}} K^{-1}_{j} \text{sgn} (x - x^{\prime})$,
where the elements of the $K$ matrix are $K_{1} = 5, K_{2} = -5, K_{3} = 3$. 
These bare edge modes may undergo subsequent renormalization due to disorder-induced tunneling 
and intermode interactions. Our variational analysis indicates that for 
sufficiently smooth potentials, the outermost downstream $e/5$ mode would be located 
far from the inner two modes and hence, couple very weakly with those. This motivates the idealization 
that $\phi_{1}$ is completely decoupled from $\phi_{2,3}$ (cf.~Fig.~3). 
The inner modes correspond to an upstream $e/5$ mode ($\phi_{2}$) and a downstream $e/3$ mode ($\phi_{3}$). 
Because of the unequal charges, this pair of counterpropagating modes cannot be localized by disorder-induced backscattering. Instead, 
the renormalized modes support excitations with generic (nonuniversal) charges $e_{u}$ and 
$e_{d}$~\cite{Yuval_AP_2017}, where $u$ and $d$ denote the upstream and downstream direction, respectively. 
Interestingly, under certain conditions, the upstream mode may be charge neutral, i.e., $e_u = 0$. 
In this case, the bulk-boundary correspondence dictates that $e_{d} = 2e/15$. 
The emergent edge structure thus consists of two downstream modes $\phi_{1}$ (with charge $e/5$) and 
$\phi_c$ (charge $2e/15$) and one upstream neutral mode $\phi_{n}$. 

The emergent $\phi_{n}$ mode has several experimental consequences. It may lead to an
upstream heat flow without an accompanying charge current. Such observations, consistent with either charge equilibration or 
the presence of a coherent neutral mode, were reported in Refs.~\cite{Inoue2014,Yacoby2012}. 
Another major consequence is that neutral modes may suppress 
the visibility of anyonic interference in electronic Mach-Zehnder setups~\cite{Moshe_PRL_2016}, as  was reported in
Ref.~\cite{Heiblum2019}. Finally, consider the QPC setup discussed previously, which may be tuned 
to a quantized conductance plateau of $1/5$. Because of the presence of counterpropagating modes
(which are fully reflected at the QPC), the system may exhibit shot noise even though the conductance 
is quantized; the ensuing Fano factor may also be quantized if certain conditions are satisfied~\cite{Spanslatt2020,Jinhong2020}.

\textit{Conclusions.}--Transport measurements~\cite{Heiblum2019,Inoue2014} suggest that the orthodox edge 
models~\cite{Wen1990,Wen1992,WenBook} do not hold even for relatively simple QH states. 
Specifically, experiments point to the presence of upstream mode(s) at the edge of a bulk $1/3$ state. 
Motivated by this surprising finding, here we study the edge structure of the
$\nu = 1/3$ Laughlin state as a function of the slope of the boundary confining potential. 
We find that an additional incompressible side strip is nucleated for sufficiently smooth potentials. 
Such a configuration allows the coarse-grained electronic density to follow the 
confining potential, while at the same time facilitating the formation of gapped QH states locally. 
Our analysis reveals that the filling factor of this side strip depends on the slope of the confining potential. 
For a moderate slope, a $\nu = 1/3$ side strip arises while for a sufficiently small slope  
the side strip is described by $\nu = 1/5$. The latter structure supports three gapless chirals: an $e/3$ downstream 
mode and a counterpropagating pair of $e/5$ modes. Subsequent renormalization, driven by intermode interactions and 
disorder-induced-tunneling among the downstream $e/3$ and upstream $e/5$ modes, may lead to the emergence of an upstream 
neutral mode, and may account for the observations of Refs.~\cite{Heiblum2019,Inoue2014}. 
We also discuss additional experimental manifestations for the reconstructed edge structures. 
We expect that edge stripes with more complex structure may arise upon fine-tuning the interplay 
between interaction and confining potential. Detailed investigations along these lines, including in 
engineered geometries~\cite{Design1,Design2}, is left to future work. 

\begin{acknowledgments}
We acknowledge useful discussions with Moty Heiblum.
U. K. was supported by the Raymond and Beverly Sackler Faculty of Exact Sciences at Tel Aviv University
and by the Raymond and Beverly Sackler Center for Computational Molecular and Material Science.
M.G. was supported by the US-Israel Binational Science Foundation
(Grant No.~2016224). Y.G. was supported by CRC~183 (project~C01), the Minerva Foundation,
DFG Grant No.~RO~2247/11-1, MI~658/10-2,
the German Israeli Foundation (Grant No.~I-118-303.1-2018), and the Helmholtz International Fellow Award.
\end{acknowledgments}

\onecolumngrid
\clearpage

\setcounter{affil}{0}
\setcounter{page}{1}
\renewcommand{\thefigure}{S\arabic{figure}}
\setcounter{figure}{0}
\renewcommand{\theequation}{S\arabic{equation}}
\setcounter{equation}{0}
\renewcommand\thesection{S\arabic{section}}
\setcounter{section}{0}

\title{Supplemental Material for ``Emergence of Neutral Modes in Laughlin-like Fractional Quantum Hall Phases''}
\author{Udit Khanna}
\affiliation{Raymond and Beverly Sackler School of Physics and Astronomy, Tel-Aviv University, Tel Aviv, 6997801, Israel}
\affiliation{Department of Condensed Matter Physics, Weizmann Institute of Science, Rehovot 76100, Israel}

\author{Moshe Goldstein}
\affiliation{Raymond and Beverly Sackler School of Physics and Astronomy, Tel-Aviv University, Tel Aviv, 6997801, Israel}
\author{Yuval Gefen}
\affiliation{Department of Condensed Matter Physics, Weizmann Institute of Science, Rehovot 76100, Israel}

\begin{abstract}
This supplemental material provides additional details regarding our numerical analysis. 
\end{abstract}

\maketitle

\begin{centering}
\subsection*{Basic Setup}
\end{centering}

We employ the disk geometry to analyze the edge of the $\nu = 1/3$ state and use a rotationally symmetric gauge, 
$e \vec{A} /\hbar = (-y/2\ell^{2}, x/2\ell^{2})$, where $\ell = \sqrt{\hbar/eB}$ is the magnetic length. 
Due to the rotational symmetry of the system, the single-particle states may be labelled by eigenvalues 
of the angular momentum ($\hat{L}$). We denote the states in the lowest Landau level (LLL) as $\phi_{m}$ with $m = 0, 1, 2, \dots$. 
The corresponding wavefunction is $\phi_{m} (\vec{r}\,) 
= \left( r / \ell \right)^{m} e^{-im\theta_{\text{r}}} e^{-\left(\frac{r}{2\ell}\right)^2} / \sqrt{2^{m+1} \pi m! \ell^2} $,
where $re^{-i \theta_{\text{r}}} = x - iy$ is the position of the electron. The state $\phi_{m}$ is 
strongly localized around $r = \sqrt{2 m} \ell$ and has angular momentum $\hbar m$. 

Restricting the Hilbert space to LLL and ignoring electronic spin, the Hamiltonian is $H = H_{\text{ee}} + H_{\text{c}}$,
where $H_{\text{ee}}$ is the electronic interaction and $H_{\text{c}}$ is the edge confining potential 
(assumed to be rotationally symmetric). 
Note that $H$ commutes with $\hat{L}$. Therefore, the total angular momentum may be used to
label the many-body states in our analysis. 
Defining $E_{c} = e^2/\epsilon_0 \ell$ as the Coulomb energy scale and $c_{m}$ as the annihilation operator
corresponding to $\phi_{m}$, we have, 
\begin{align} 
  H_{\text{ee}} &= \frac{E_{\text{c}}}{2} \sum_{i \neq j} \frac{\ell}{|\vec{r}_{i} - \vec{r}_{j}|} \\
  &\equiv \frac{E_{\text{c}}}{2} \sum_{m_{1},m_{2},n} V_{m_1 m_2 ; n}^{ee} 
c_{m_1 + n}^{\dagger} c_{m_2}^{\dagger} c_{m_2 + n} c_{m_1}, \\
  H_{\text{c}} &= \sum_{m} V_{m}^{\text{c}} \, \, c_{m}^{\dagger} c_{m}. \label{eq:S2}
\end{align}
To model the edge potential, we consider a uniformly charged disk with radius $R$ 
and positive charge density $\sigma$, located a distance $d$ away from the electronic plane~\cite{SKunYangIQHS,SKunYang_2002,SKunYang_2003}.
The electrostatic potential of this disk on the electrons is given by,
\begin{align}
  V_{c} (r) = \int_{0}^{R} dr^{\prime} \int_{0}^{2\pi} d \theta \frac{E_{c} \sigma}{\sqrt{d^2 + r^2 + {r^{\prime}}^2 - 
  2 r^{\prime} r \cos \theta}}. 
\end{align}
$V_{m}^{\text{c}}$ in Eq.~(\ref{eq:S2}) are the matrix elements of $V_{c}(r)$. 
We use $\sigma = (1/3) / 2\pi \ell^2$ and $R^2 = 2(N_{S} + N_{B}) \ell^2$, in order to maintain overall charge neutrality. 
For $d = 0$ this confining potential is very sharp, and it get shallower as $d$ increases.

\begin{centering}
\subsection*{Variational Analysis}
\end{centering}

Figure~1 depicts the two classes of variational states considered in this work. Both classes
represent the product state of a bulk $\nu = 1/3$ state with an annulus of the $\nu = 1/m_{S}$ state ($m_{S} = 3, 5$). 
The bulk state (denoted as $|\psi_{\frac{1}{3}, N_{B}}\rangle$) comprises $N_{B}$ electrons and is 
well described by the (unnormalized) Laughlin wavefunction~\cite{SLaughlin83}
\begin{align} \label{eq:S5}
  \Psi_{\frac{1}{3}, N_{B}} = \prod_{i=1}^{N_{B}} \bigg[ \prod_{j > i} \big( z_i - z_j \big)^{3} \bigg] e^{-\frac{1}{4} \sum_{i} |z_{i}|^2}. 
\end{align}
Here $z_{j} = (x_{j} - i y_{j})/\ell$ is coordinate of the $j^{\text{th}}$ electron. The annulus state
(denoted as $|\psi_{\frac{1}{m_{S}}, N_{S}, M_{S}}\rangle$) is a $\nu = 1/m_{S}$ Laughlin state consisting of 
$N_{S}$ electrons and $M_{S}$ quasiholes at the origin. The corresponding wavefunction may be written as,
\begin{align} \label{eq:S6}
  \Psi_{\frac{1}{m_{S}}, N_{S}, M_{S}} = \prod_{i=1}^{N_{S}} \bigg[ z_{i}^{M_{S}} \bigg] \bigg[ \prod_{j > i} 
  \big( z_i - z_j \big)^{m_{S}} \bigg] e^{-\frac{1}{4} \sum_{i} |z_{i}|^2}. 
\end{align}
Note that the single-particle state with smallest angular momentum which may be occupied in the annulus state is $\phi_{M_{S}}$. 
Then if $L_{S}$ denotes the number of guiding centers between the bulk and annulus state, we have $M_{S} = 3N_{B} - 2 + L_{S}$. 

The angular momentum (in units of $\hbar$) of the ($\nu = 1/m$) Laughlin state with $N$ particles is $\frac{m}{2} N (N-1)$. 
Introducing $M$ quasiholes at the origin increases the angular momentum by $NM$. 
Then the total angular momentum of a product state with a $\nu = 1/m_{S}$ side strip is
\begin{align} \nonumber
  \frac{3}{2} &(N_{B}+N_{S}) (N_{B} + N_{S} - 1) \\ &+ \frac{m_{S} - 3}{2} N_{S} (N_{S} - 1) + N_{S} (L_{S} - 2).
\end{align}
The first term above is the angular momentum of the unreconstructed state. This indicates that the variational 
states may have angular momentum smaller than that of the unreconstructed state if $L_{S}$ is sufficiently
small. However, compressing a Laughlin state would increase the Coulomb repulsion and therefore 
the energy of such states is unlikely to be lower than that of the unreconstructed state. 

The energy ($\langle H \rangle$) of the product states is the sum of the energy of the individual components and
their mutual interaction energy. The energy of individual components is,
\begin{align}
  \frac{1}{\int \prod_{i} d^2 r_i \big| \Psi \big|^2} &\int \prod_{i} d^2 r_i
  \big| \Psi \big|^2 \bigg[ \sum_{i < j} \frac{E_{c} \ell}{|\vec{r}_i - \vec{r}_j|} \bigg] \nonumber \\ &+
  E_{c} \sum_{m} \langle c_{m}^{\dagger} c_{m} \rangle_{\Psi} V_{m}^{c}.
\end{align}
where $\langle c_{m}^{\dagger} c_{m} \rangle_{\Psi}$ is the occupation of the single-particle states in each
component. The mutual interaction energy of the bulk ($\Psi_{B}$) and strip ($\Psi_{S}$) components is,
\begin{align}
  E_{c} \sum_{i \in \text{B}} \sum_{j \in \text{S}}
  \langle c_{i}^{\dagger} c_{i} \rangle_{\Psi_{\text{B}}}
  \langle c_{j}^{\dagger} c_{j} \rangle_{\Psi_{\text{S}}}
  \bigg( V_{i j ; 0}^{ee} - V_{i i ; j-i}^{ee} \bigg).
\end{align}
The Coulomb energy of each component maybe evaluated using the standard Metropolis algorithm~\cite{SMetropolis53,SLaughlin83,SJainCF}.
The average occupation may be evaluated similarly~\cite{SMacDonald93,SIQHS2020}.

\end{document}